\documentclass[twocolumn,preprintnumbers,amsmath,amssymb,aps,prb,floatfix,groupedaddress]{revtex4}
\usepackage{graphicx}
\usepackage{dcolumn}
\usepackage{bm}

\usepackage{bm} 
\usepackage{epsfig}
\usepackage{color}
\usepackage{multirow}
\usepackage{amsmath}
\begin{document}

\title{Anderson localizaion for semi-Dirac semi-Weyl semi-metal}

\author{Swapnonil Banerjee}

\date{\today}

\begin{abstract}
In recent years ihe semi-Dirac semi-Weyl semi-metal has been of interest due to its
naturally occurring point Fermi surface and the exotic anisotropic band-structure near
the Fermi surface, which is linear (graphene-like) in one direction of the Brillouin
zone, but quadratic in a direction perpendicular to it. In this paper the effect of a
magnetic adatom impurity in a semi-Dirac system is
studied. As in a metal, the magnetic impurity in a semi-Dirac system interacts with the sea of
conduction electrons and gives rise to magnetism. The transition of the semi-Dirac system
from the non-magnetic to the magnetic phase is studied as a function of
the impurity energy, the strength of hybridization between the impurity and the bath as well as
that of the electron electron interaction at the impurity atom.
The results are compared and contrasted with those of graphene and ordinary metal.
Since the semi-Dirac and the Dirac dispersion share similar features,e.g, both are particle hole
symmetric and linear in one direction, the two systems share resemblances in
their characteristics in the presence of a magnetic impurity. But some
features are unique to the semi-Dirac dispersion.
\end{abstract}

\maketitle

\section{Introduction}
It has been about about a decade since a graphene layer has been isolated in the laboratory successfully.
Its unique linear (massless Dirac, properly called Weyl)low energy band structure has been studied extensively.\cite{graphene_nat05,antonio}
One of the low energy features of graphene is the point Fermi surface aspect. In the graphene bandstructure, there is gap
thoroughout the Brillouin zone except for at the point where the bands touch each other. The Fermi
(E$_F$) lies exactly at that point. Away from the Fermi surface the bands open up in a linear manner for as long
as the energy is not too large. Another new point Fermi surface system, named `semi-Dirac' or `semi-Weyl' has been of interest in recent years.
It has the aspects of both the conventional and the unconventional dispersions.
The dispersion is linear (``massless'', Dirac-Weyl) in one of the directions of the
two-dimensional (2D) layer, and is conventional quadratic (``massive'')
in the perpendicular direction. At directions between the axes the
dispersion is intermediate and highly direction-dependent.
Volovik obtained such a spectrum at the point node in the A-phase of superfluid $^3$He~[\onlinecite{volovik1}]
and studied its topological robustness.\cite{volovik2}  
Montambaux's group discovered the semi-Dirac spectrum in a model based upon the honeycomb lattice geometry,
which also happens to be crystal geometry for graphene.\cite{Montambaux_One} 
The model has a broken rotational symmetry such that hopping to two nearest neighbors is $t$ but to
the third neighbor is $t'$.  When $t'$ differs from $t$, the graphene ``Dirac points''
move away from the $K$ and $K'$ points, and for the specific choice $t' = 2t$ they merge, resulting in
the semi-Dirac spectrum.  This group began a study of low energy properties of such a
system,\cite{Montambaux_One} which was continued by Banerjee {\it et al.}\cite{Tightbinding_SemiDirac}.
The point Fermi surface along with the anisotropic dispersion (massive in one direction, massless
in the orthogonal direction, and hybrid of the two in between)
characterizes a semi-Dirac dispersion in an unique manner. 
The semi-Dirac dispersion has recently been experimentally realized in the context of optical lattices\cite{ColdAtom_Sd}.
There ultracold atoms, which mimic the electrons in a real condensed matter
sustem, have been studied in an adjustable honeycomb lattice environment. By tuning the hopping parameters to the
right anisotropy the merging of the Dirac points into a single semi-Dirac point has been achieved.
From the simulation standpoint Pardo and Pickett\cite{PardoSd1,PardoSd2} reported the finding
of the semi-Dirac dispersion in ultrathin (001) VO$_2$ layers embedded in TiO$_2$ using Density functional theory calculations.

In this paper we investigate how a single magnetic impurity hybridizes with the conduction electrons in
a semi-Dirac system to give rise to the magnetic non-magnetic phase transition. This
study has previously been conducted in the context of an ordinary metal\cite{Anderson_Imp} and graphene\cite{Grphn_And_Imp}.
A metal with a magnetic impurity in it does not show any magnetism if the energy of the
impurity atom is larger than the Fermi energy. But a somewhat unusual non-magnetic to magnetic phase
transition takes place when a magnetic impurity is embedded in the graphene crystal.
Due to the anomalous broadening of the impurity density of states, magnetism
can be observed in graphene even when the energy of the impurity atom is larger than the Fermi
energy. In this paper it has been investigated how this issue as well as various other ones
play out in the context of the semi-Dirac system with a magnetic impurity in it.
The complete Hamiltonian for this problem consists of a 2 band term\cite{Montambaux_One}
representing the semi-Dirac conduction electrons,
the onsite energy term of the localized impurity state, the term representing
the hybridization of the impurity with the conduction electrons and finally the term
representing the Coulomb interaction between the electrons at the impurity site. In this paper it is
investigated how the semi-Dirac system makes a transition from a magnetic state to a non-magnetic one
and vice versa as parameters like the impurity energy $\epsilon_0$, the strength of the electron electron Coulomb interaction $U$,
and the the impurity conduction electron hybridization parameter $V$ are allowed to scan the parameter space.


\section{The semi-Dirac system with impurity}
 In this section we will discuss the four major components, as mentioned in the previous section, of the Hamiltonian for the semi-Dirac system with a magnetic impurity in details. Those are denoted by $H_{TB}, H_V, H_f$, and $H_U$ respectively. Mathematical expressions for each of the above terms are given below.
 $H_{TB}$, the two band tight binding part of the Hamiltonian, which corresponds to the semi-Dirac conduction electrons, is written as
\begin{eqnarray}
H_{TB}=\sum_{k,\sigma} V_{12}(k)c_{1,\sigma}^\dag(k)c_{2,\sigma}(k)+ h.c,
\label{eq:semiDirac_TB}
\end{eqnarray}
where $c_{1(2),\sigma}^\dag(k)$, and $c_{1(2),\sigma}(k)$ are the creation and the destruction operators of the conduction electron with the spin index $\sigma$; $1$ and $2$ being the band indices. The off-diagonal element $V_{12}(k)$ is given by

\begin{eqnarray}
V_{12}(k)=vk_x-i\frac{k_y^2}{2m},
\label{eq:V_sD}
\end{eqnarray}
$v$, and $m$ are the characteristic velocity and the mass parameters of a semi-Dirac system ($\hbar$ has been set to $1$).
Eq.~\ref{eq:semiDirac_TB} is the second quantized, spin added version of the single particle, spinless semi-Dirac Hamiltonian appearing in [\onlinecite{Montambaux_One}].  The part of the Hamiltonian that represents the hybridization between the conduction electrons and the impurity energy state, denoted as $H_{V}$, is given by \cite{Grphn_And_Imp}
\begin{eqnarray}
H_{V}=\frac{V}{\sqrt{N}}\sum_{k,\sigma}(f_{\sigma,k}^\dag c_{1,\sigma}(k)+f_{\sigma,k}^\dag c_{2,\sigma}(k)+h.c),
\label{eq:Band_Imp_cpl}
\end{eqnarray}
where $f_{\sigma,k}^\dag$, and $f_{\sigma,k}$ are the creation and the destruction operators corresponding to the impurity state of  spin $\sigma$. $N$ denotes the normalization factor
The onsite energy term for the impurity is given by
\begin{eqnarray}
H_f=\epsilon_0\sum_{k,\sigma}f^\dag_{\sigma,k} f_{\sigma,k}.
\label{eq:ImpOnsite}
\end{eqnarray}
$\epsilon_0$ is the energy of the localized impurity state. So far although the spin indices appear in the Hamiltonian, operators with opposite spins are decoupled from each other. The introduction of the Coulomb interaction term $H_U$, the real space expression of which is given in the following changes that.
\begin{eqnarray}
H_U=Uf^\dag_\uparrow f_\uparrow f^\dag_\downarrow f_\downarrow,
\label{eq:ImpCoul}
\end{eqnarray}
It is clearly seen from the above equation that operators with opposite spins interact with each other. $U$ appearing in the expression for $H_U$ is a constant, which denotes the strength of the Coulomb repulsion between the electrons of opposite spins at the impurity site.
The complete Hamiltonian (H)is obtained by adding together all the individual pieces ($H_{TB}, H_V, H_f$, and $H_U$), that have been introduced so far.

\begin{eqnarray}
H=H_{TB}+ H_V + H_f + H_U
\label{eq:NetH}
\end{eqnarray}
Due to the presence of the Coulomb interaction term $H_U$, the Hamiltonian in Eq.~\ref{eq:NetH} can't be solved exactly. In order to render $H_U$ more tractable we resort to the mean field technique. The mean field decomposition of $H_U$ is given as follows.
\begin{eqnarray}
H_U \approx U[f^\dag_\uparrow f_\uparrow \langle f^\dag_\downarrow f_\downarrow \rangle +
\langle f^\dag_\uparrow f_\uparrow \rangle f^\dag_\downarrow f_\downarrow
-\langle f^\dag_\uparrow f_\uparrow \rangle \langle f^\dag_\downarrow f_\downarrow \rangle]
\label{eq:Coul_MeanField}
\end{eqnarray}
The last term in the above equation is a constant and can safely be ignored.
The first two terms have a common trait: both of the them have the number operator of the impurity
state of a given type of spin, multiplied by a scalar which is the average of the number
operator of the opposite spin. Such a term when grouped with the on-site energy term
for the same spin specie appearing in $H_f$ produces a scalar like $\epsilon_\sigma$ defined as:
$\epsilon_\sigma \equiv \epsilon_0 + Un_{-\sigma}$, where $n_{\sigma}$ denotes the average
of the number operator of spin $\sigma$. $-\sigma$ indicates the spin type opposite to that indicated by $\sigma$.
Hence under the mean field approximation, adding Eq.~\ref{eq:ImpOnsite} and Eq.~\ref{eq:Coul_MeanField} together
we obtain
\begin{eqnarray}
H^\prime_f=\sum_{k,\sigma}\epsilon_\sigma f^\dag_{\sigma,k} f_{\sigma,k}.
\label{eq:ImpMeanFinal}
\end{eqnarray}
The final mean field expression of the Hamiltonian is obtained by adding together Eqs.~\ref{eq:semiDirac_TB}, ~\ref{eq:Band_Imp_cpl}, and ~\ref{eq:ImpMeanFinal}
as follows
\begin{eqnarray}
H_{MF}=H_{TB}+H_V+H^\prime_f,
\label{eq:MeanFinal}
\end{eqnarray}
Instead of using the exact Hamiltonian given by Eq.~\ref{eq:NetH}, the simplified mean field version of it as given by the above equation will be used for all the calculations in this paper.

\section{Green's function for the impurity state}
 In order to find out if the semi-Dirac system is in the magnetic phase or not for a certain parameter set, one needs to
solve for impurity occupation numbers corresponding to two different spins in a self consistent manner. If the difference
between the occupation numbers is non-zero that would imply that the system is magnetic phase, whereas a zero difference
between them would be indicative of a non-magnetic phase.
In order to determine impurity occupation numbers finding the Green's function for the impurity state is essential. From the
Green's function the impurity density of states can be obtained, which upon integration will give an expression for
the impurity occupation number. The impurity Green's function $G_{ff,\sigma}^R$ for the semi-Dirac system is defined as
$G_{ff,\sigma}^R \equiv <T[f_\sigma f^\dag_\sigma]>$,$T$ denoting the time ordering operator.
An expression for $G_{ff,\sigma}^R$ derived using the above definition is given by\cite{Grphn_And_Imp}
\begin{eqnarray}
G_{ff,\sigma}^R  =  \frac{1}{\omega-\epsilon_\sigma-\Sigma_{ff, \sigma}^R-i0^\dag}.
\label{eq:GreensFunction}
\end{eqnarray}
$\Sigma_{ff,\sigma}^R$ in the above equation is the self energy is written in terms of non-interacting Green's function as follows
\begin{eqnarray}
\Sigma_{ff, \sigma}^R  =  \frac{V^2}{N_b}\sum_k G_{cc,\sigma}^R,
\label{eq:SelfEnergy_One}
\end{eqnarray}
The non-interacting Green's function $G_{cc,\sigma}^R$ in the above equation is given by
\begin{eqnarray}
G_{cc,\sigma}^R  =  \sum_k \frac{\omega}{\omega^2-\varepsilon^2+i0^\dag \text{sign}(\omega)},
\label{eq:Green_NonInt}
\end{eqnarray}
where $\varepsilon^2$ is the square of the quasi-particle energy of the semi-Dirac system, which
in the momentum space takes the form $\varepsilon^2=v^2k_x^2+\frac{k_y^4}{4m^2}$.
The summation over momentum in Eq.~\ref{eq:SelfEnergy_One} is converted into an integral involving
the density of states of the semi-Dirac quasi-particle. The integration is carried out
from $-D$ to $D$, where $D$ is the cut off energy. The Semi-Dirac
dispersion is assumed to remain valid for $|\varepsilon|< D$. It turns out that the integral can be evaluated
in closed form. From Eqs.~\ref{eq:SelfEnergy_One} and ~\ref{eq:Green_NonInt} one finally obtains

\begin{eqnarray}
\Sigma_{ff,\sigma}^R  =  -\frac{3V^2}{2D^{3/2}}\left
[\omega I(\omega)+i\frac{\pi}{2}|\omega|\Theta(D-|\omega|)\right],
\label{eq:SelfEnergy_Two}
\end{eqnarray}

where $I(\omega)$ is given by
\begin{eqnarray}
I(\omega)=\frac{1}{2\sqrt{|\omega|}}\ln \left \lvert \frac{\sqrt{D}-\sqrt{|\omega|}}{\sqrt{D}+\sqrt{|\omega|}} \right \rvert +\frac{1}{\sqrt{|\omega|}}\arctan{\frac{\sqrt{D}}{\sqrt{|\omega|}}}
\label{eq:SelfEnergyFactor}
\end{eqnarray}

At this point a brief discussion
about the choice of $D$ is warranted. D is ascertained following the Debye
prescription\cite{Grphn_And_Imp}. According to the Debye prescription, D is determined based
on the conservation of number of states in the Brillouin zone. $D$ for the semi-Dirac
system is a function of the
adjustable parameters $m$ and $v$. Since $m$ and $v$ need to be determined empirically
from the real physical system, $D$ can differ from one semi-Dirac system to another.
To be able to compare the results with graphene, the value of $D$ for the semi-Dirac is
chosen to be the same as that in graphene, for which $D \sim 7$ eV\cite{Grphn_And_Imp}.
[The semi-Dirac parameters $m$ and $v$ depend on the
tightbinding parameter $t$, and the bond-length $a$ \cite{Montambaux_One}. Using the same numerical values for
$t$ and $a$ as are used for graphene, $D$ for
semi-Dirac has been verified to be of the same order of
magnitude as that in case of graphene. Hence the aforementioned choice of $D$ is reasonable.] As long as the Fermi
energy $\mu$ is much less compared to $D$, the physics does not get affected
by a specific choice of the value of $D$.

The localized density of states for the impurity atom is obtained from the imaginary part
of the Green's function \cite{Grphn_And_Imp}.
\begin{eqnarray}
\rho_{ff}(\omega,\sigma)  =  -Im(\frac{1}{\pi}G_{ff,\sigma}^R).
\label{eq:DOSImp_One}
\end{eqnarray}

From Eqs. ~\ref{eq:GreensFunction}, ~\ref{eq:SelfEnergy_Two} and ~\ref{eq:DOSImp_One} we obtain
\begin{eqnarray}
\rho_{ff,\sigma}(\omega)  =  \frac{1}{\pi}\frac{\frac{3\pi}{4}\frac{V^2}{D^{3/2}}|\omega|^{1/2}\Theta(D-|\omega|)}
                       {(\omega Z^{-1}(\omega)-\epsilon_\sigma)^2+(\frac{3\pi}{4}\frac{V^2}{D^{3/2}})^2|\omega|},
\label{eq:DOSImp_Two}
\end{eqnarray}

where $Z^{-1}(\omega)$ is given by
\begin{eqnarray}
Z^{-1}(\omega)=1+\frac{3V^2}{2D^{3/2}}I(\omega).
\label{eq:ZInv}
\end{eqnarray}
The impurity density of states appearing in Eq.~\ref{eq:DOSImp_Two}
looks very different when compared to the Lorentzian impurity density of states for an ordinary metal\cite{Anderson_Imp}.
It also differs from the impurity density of states of graphene, e.g, in the presence of a square root of the absolute value of
the energy $\omega$ instead of simply the absolute value of $\omega$ in the numerator.
The occupation $(n_\sigma)$ of the impurity is given as the integral of $\rho_{ff,\sigma}$ up to the Fermi energy as follows
\begin{eqnarray}
n_\sigma=\int_{-\infty}^\mu d\omega \rho_{ff}(\omega),
\label{eq:nsigma_One}
\end{eqnarray}
The above equation is written as the sum of two separate integrals over the energy ranges $(-\infty \text{ to } 0)$
and $(0 \text{ to } \mu)$ respectively. Under the assumption that the semi-Dirac dispersion is valid from $-D$ to $D$,
as was explained before, in one of the integrals $-\infty$ should be replaced by $-D$.
$\mu$ being much smaller than $D$, the upper limit of the integral remains unchanged.
from Eqs.~\ref{eq:DOSImp_Two} and ~\ref{eq:nsigma_One}, we obtain
\begin{eqnarray}\label{eq:nsigma_Final}
n_\sigma=I_1+I_2,
\end{eqnarray}

where
\begin{subequations}\label{eq:IOneANDITwo}
\begin{eqnarray}\label{eq:IOne}
 I_1=\frac{\Delta}{\pi}\int_{-D}^0 d\omega \frac{(-\omega)^{1/2}}{(\omega Z^{-1}(\omega)-\epsilon_{\sigma})^2-\Delta^2\omega},
\end{eqnarray}
\begin{eqnarray}\label{eq:ITwo}
 I_2=\frac{\Delta}{\pi}\int_{0}^\mu d\omega \frac{\omega^{1/2}}{(\omega Z^{-1}(\omega)-\epsilon_{\sigma})^2+\Delta^2\omega},
\end{eqnarray}
\end{subequations}
where $\Delta\equiv(3\pi/4)(V^2/D^{\frac{3}{2}})$. $n_\sigma$ appearing in the left side of Eq.~\ref{eq:nsigma_Final}
depends on $n_{-\sigma}$ through the term $\epsilon_{\sigma}$ appearing on the right side of the same equation. Hence Eq.~\ref{eq:nsigma_Final} needs to be solved self consistently.
For a given $\omega$ the integrands in the above equations are large when $Z^{-1}(\omega)\omega \approx \epsilon_\sigma$.
Referring to Eq.~\ref{eq:ZInv} and keeping only the dominant term in the expression of $Z^{-1}(\omega)\omega$ for a small $\Delta$ and not too small $\omega$, it is clear that $\omega \sim \epsilon_\sigma$.
Hence $Z^{-1}(\omega)$ can be approximated by $Z^{-1}_\sigma\equiv Z^{-1}(\epsilon_{\sigma})$. With this approximation $I_1$ and $I_2$ can be evaluated in closed forms. It has been checked that the results obtained with the approximation match rather well with those using the complete $\omega$ dependence of $Z^{-1}(\omega)$. To keep the final closed form expressions of the integrals neat, the following quantities
are defined.

\begin{subequations}
\begin{eqnarray}\label{eq:const_1}
 a \equiv \Delta^2+2Z_\sigma^{-1}\epsilon_\sigma,
\end{eqnarray}
\begin{eqnarray}\label{eq:const_2}
 b \equiv \Delta^2-2Z_\sigma^{-1}\epsilon_\sigma.
\end{eqnarray}
\end{subequations}

Evaluating $I_1$ and $I_2$ as given by Eq.~\ref{eq:IOneANDITwo} we obtain
\begin{subequations}\label{eq:n_approx_closed}
\begin{eqnarray}\label{eq:n_sig_minus}
 I_1 = \frac{2\Delta}{\pi Z_\sigma^{-2}}\\\nonumber \left [\frac{J_1}{\sqrt{L_1}}\arctan{\sqrt{\frac{D}{L_1}}}
 + \frac{J_2}{\sqrt{L_2}}\arctan{\sqrt{\frac{D}{L_2}}}\right ]
\end{eqnarray}
\begin{eqnarray}\label{eq:n_sig_plus}
 I_2 = \frac{2\Delta}{\pi Z_\sigma^{-2}}\\\nonumber \left [\frac{J_1^\prime}{\sqrt{L_1^\prime}}\arctan{\sqrt{\frac{\mu}{L_1^\prime}}}
 + \frac{J_2^\prime}{\sqrt{L_2^\prime}}\arctan{\sqrt{\frac{\mu}{L_2^\prime}}}\right ]
\end{eqnarray}
\end{subequations}

where
\begin{subequations}
\begin{eqnarray}\label{eq:J_1}
 J_1 & = &  \frac{1}{2}\left[ 1+\frac{a}
 {\sqrt{a^2-4Z_\sigma^{-2}\epsilon_\sigma^2}}\right].
\end{eqnarray}
\begin{eqnarray}\label{eq:J_2}
 J_2 & = & \frac{1}{2}\left[ 1-\frac{a}
 {\sqrt{a^2-4Z_\sigma^{-2}\epsilon_\sigma^2}}\right].
\end{eqnarray}
\begin{eqnarray}\label{eq:L_1}
 L_1 & = & \frac{1}{2Z^{-2}(\epsilon_\sigma)}[a
           +\sqrt{a^2-4Z_\sigma^{-2}\epsilon_\sigma^2}].
\end{eqnarray}
\begin{eqnarray}\label{eq:L_2}
 L_2 & = & \frac{1}{2Z^{-2}(\epsilon_\sigma)}[a
 -\sqrt{a^2-4Z_\sigma^{-2}\epsilon_\sigma^2}].
\end{eqnarray}
\begin{eqnarray}\label{eq:J_1Prime}
 J_1^\prime & = & \frac{1}{2}\left[ 1+\frac{b}
 {\sqrt{b^2-4Z_\sigma^{-2}\epsilon_\sigma^2}}\right].
\end{eqnarray}
\begin{eqnarray}\label{eq:J_2Prime}
 J_2^\prime & = & \frac{1}{2}\left[ 1-\frac{b}
 {\sqrt{b^2-4Z_\sigma^{-2}\epsilon_\sigma^2}}\right].
\end{eqnarray}
\begin{eqnarray}\label{eq:L_1Prime}
 L_1^\prime & = & \frac{1}{2Z^{-2}(\epsilon_\sigma)}[b
 +\sqrt{b^2-4Z_\sigma^{-2}\epsilon_\sigma^2}].
\end{eqnarray}
\begin{eqnarray}\label{eq:L_2Prime}
 L_2^\prime & = & \frac{1}{2Z^{-2}(\epsilon_\sigma)}[b
 -\sqrt{b^2-4Z_\sigma^{-2}\epsilon_\sigma^2}].
\end{eqnarray}
\end{subequations}
$n_\sigma$ is obtained self consistently from Eq.~\ref{eq:nsigma_Final} using the Matlab minimization routine `fsolve'. A shift in the
difference between the values of $n_\sigma$ for two different $\sigma$'s corresponding to two different
spins, from the zero to a non-zero value implies that the the system has made a transition from the
non-magnetic to the magnetic phase. Defining the variables $x$, and $y$ as
$x \equiv \frac{D\Delta}{U}$, and $y \equiv (\mu-\epsilon_0)/U$,
the boundary (which henceforth will simply be referred to as the transition curve) separating the
magnetic phase from the non-magnetic one is plotted as
a function of $x$ and $y$ [details to be given in the `Results' section].
The more the number of grid-points in the $x-y$ plane used to produce the
plots, the sharper is the transition curve, but that happens at the cost of
somewhat increased program runtime.
Hence an optimal number of grid points have been used to get
a fairly good idea of how the transition curve should look like for a particular parameter-set.
Then smoothing algorithm has been employed to smooth out any small irregularities that might be
present at the boundary region. But no smoothing techniques were necessary to produce
the plots for the occupation numbers for different
spins as functions of $\mu$. Both positive as well as negative values of $\epsilon_0$ have been
considered for the plots.
It turns out for the negative values of $\epsilon_0$
the convergence of the self consistent solution using fsolve optimization routine can be somewhat
poor if one works with the closed form approximations (as given by Eq.~\ref{eq:n_approx_closed}) of
the integrals appearing in Eq.~\ref{eq:nsigma_Final}. A possible reason for that
is during the optimization process $\varepsilon_\sigma$ in the expression for $Z_\sigma^{-1}$ can
go to zero resulting in making $Z_\sigma^{-1}$ a large number and thereby affecting the convergence
in a non-physical way. This does not happen when $\varepsilon_0$ is positive, since
all the terms in the expression of $\varepsilon_\sigma$ are positive and the possibility of
$\varepsilon_\sigma$ going to zero does not arise. Hence for the negative values of $\varepsilon_0$,
an un-approximated direct numerical integration of Eq.~\ref{eq:nsigma_Final} has been resorted to, for
producing the plots.

\section{Results}
\begin{figure}[ht]
\begin{center}
\includegraphics[draft=false,bb=50 225 500 550, clip, width=\columnwidth]{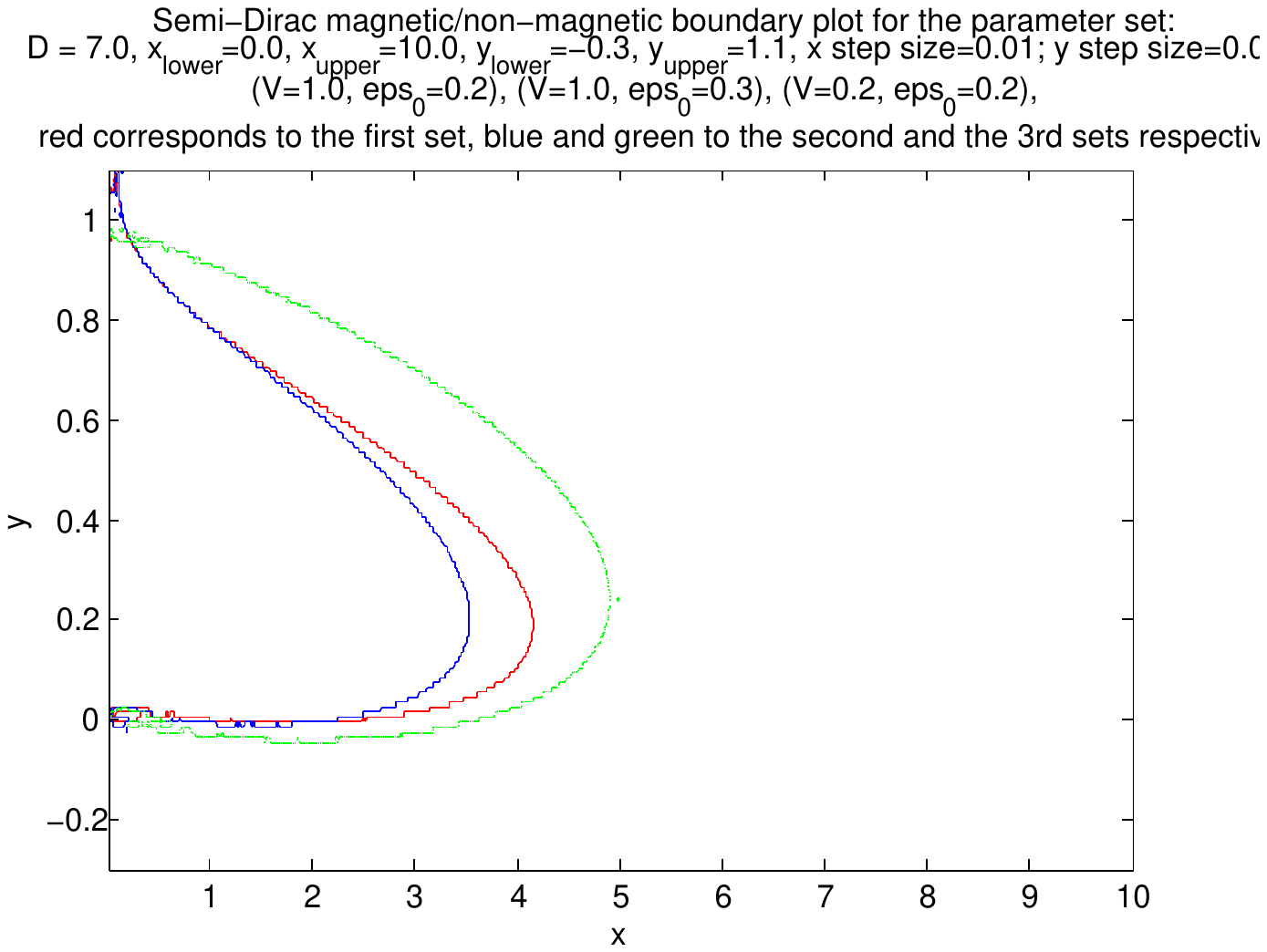}
\includegraphics[draft=false,bb=50 250 500 550, clip, width=\columnwidth]{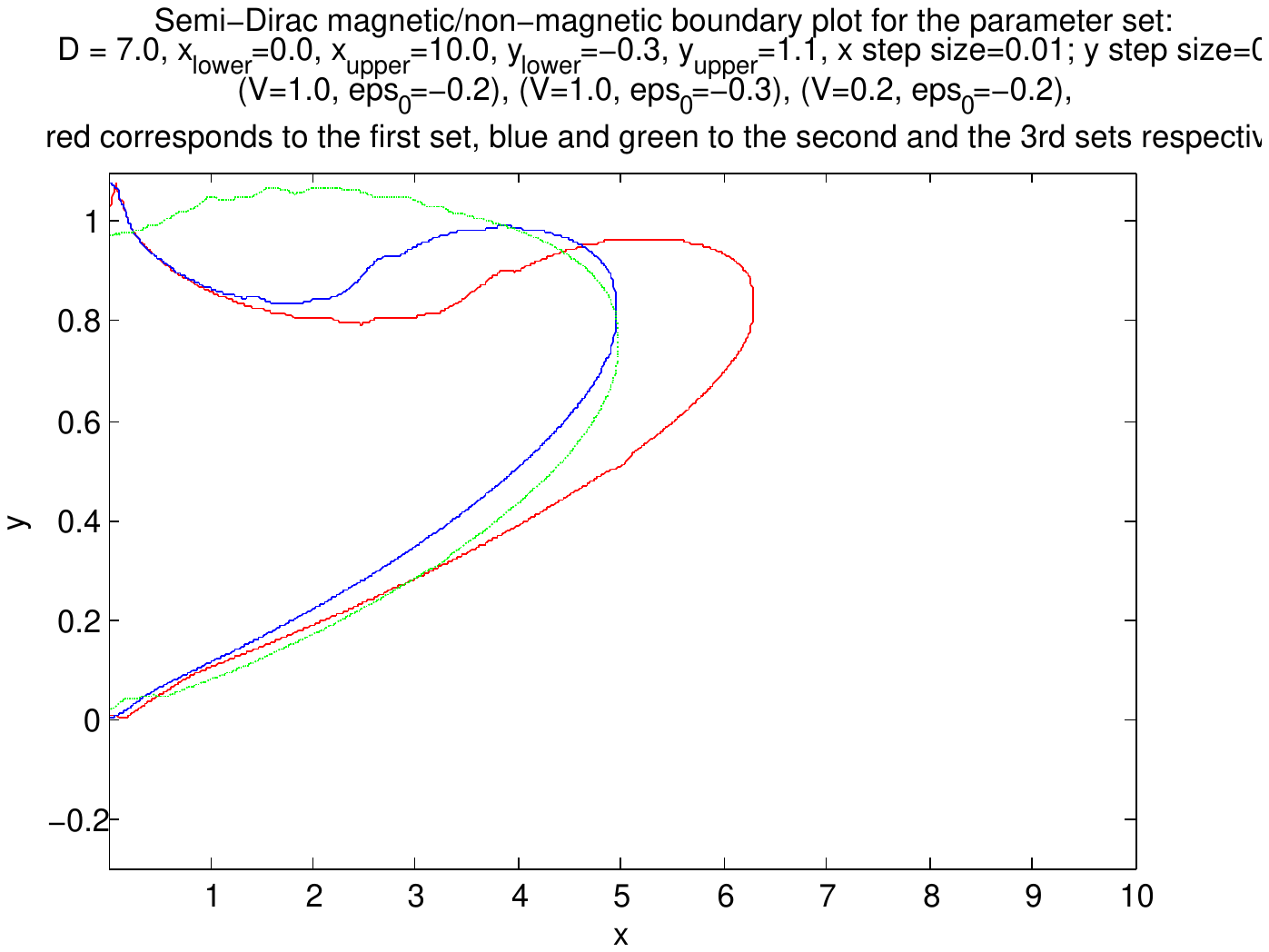}
\caption{Magnetic to non-magnetic transition curves for the following parameter sets:
Top figure: red corresponds to $\varepsilon_0=.029D, V=.14D$; blue corresponds to $\varepsilon_0=.043D, V=.14D$, and green corresponds to $\varepsilon_0=.029D, V=.03D$.
Bottom figure: red corresponds to $\varepsilon_0=-.029D, V=.14D$; blue corresponds to $\varepsilon_0=-.043D, V=.14D$, and green corresponds to $\varepsilon_0=-.029D, V=.03D$}
\label{fig:MagNMagEpsGthan0}
\end{center}
\end{figure}
In Fig.\ref{fig:MagNMagEpsGthan0} the transition curves are plotted as functions of $x$ and $y$ for positive as well as negative values of $\epsilon_0$.
As in the Dirac dispersion, they are not symmetric about $y=.5$. There is also asymmetry between $\epsilon_0$ being positive
and negative. The fact that both the semi-Dirac dispersion
and the Dirac dispersion are not symmetric about $y=.5$ can be attributed to the particle-hole symmetry breaking
due to presence of the impurity.
From the plot in the top of Fig.\ref{fig:MagNMagEpsGthan0}it is observed for the positive value of $\epsilon_0$,
the magnetic region barely crosses the line $y=0$, indicating that the system is magnetic even
when the impurity energy is slightly larger than the Fermi energy.
For an ordinary metal, the transition curve never goes below $y=0$. In case of Dirac, the magnetic region extends much further into the region where $y$ is negative with an overall negative slope, which is in direct contrast to the semi-Dirac case. When $y$ is negative, the on-site impurity energy is larger than the Fermi energy, but because of the anomalous broadening of the impurity density of states magnetism is still manifested. For the semi-Dirac the confinement of the magnetic region primarily to the positive values of $y$ can be attributed to the anomalous but somewhat weak broadening and the resulting lack of ease with which the impurity density of states crosses the Fermi energy, compared to that of the Dirac case as is explained in the following. The impurity density of states of neither the semi-Dirac nor the Dirac dispersion is a simple Lorentzian. For the semi-Dirac it falls off as $\frac{1}{|\omega|^{(3/2)}}$, as can be seen from Eq.~\ref{eq:DOSImp_Two}, whereas for the Dirac system, the tail of the impurity density of states goes as $\frac{1}{\omega}$. So, for the semi-Dirac the impurity density of states decreases faster, and hence crosses the Fermi energy with more difficulty, compared to the Dirac case. But the semi-Dirac impurity density of states does not fall off as fast as a metal, for which the density of states being Lorentzian, falls off as $\frac{1}{\omega^2}$. That explains the difference in the behavior of the transition curves in the region of negative $y$ close to $0$ between the semi-Dirac and the metal.
Whereas $y=0$ corresponds to $\epsilon_0$ being equal to $\mu$, $y=1$ corresponds to the case when the
impurity state is doubly occupied and the larger of the two impurity energy levels is equal to the Fermi energy.
From the plot in the bottom of Fig.\ref{fig:MagNMagEpsGthan0} it is observed that the transition curve for the semi-Dirac dispersion for the negative impurity energy, although somewhat similar in appearance to the Dirac, barely crosses the $y=1$ line(it extends considerably beyond $y=1$ for the Dirac case). For small $x$, the transition curve shows a kink, which implies that the curve near $x=0$ goes downward like the ordinary metal or the Dirac. This universal behavior in the transition curve is brought about by the irrelevance of the specific nature of energy momentum dispersion of the conduction electrons in the small $x$ limit, since a small value of $x$ indicates weak coupling of the magnetic impurity with the conduction electron.
A small value of $x$ corresponds to large $U$. Hence it is concluded that a large $U$ causes the system to behave universally. It turns out that compared to an ordinary metal, a much smaller value of $U$ can send the semi-Dirac system to magnetic state. For $V=1$, and $\varepsilon_0=.3$ the magnetism sets in when $U$ is as low as $.35$ (This is determined as follows. For a given $V$, and $D$, delta is calculated. Then from the top plot in Fig.\ref{fig:MagNMagEpsGthan0} the maximum value of $x$ is ascertained for which magnetism still exists. $x$ being a function of $U$, and everything else in the expression of $x$ like $D$ and $\Delta$ being known, the minimum $U$ is thus determined from the maximum possible value of $x$.) The minimum $U$ thus obtained for the semi-Dirac turns out to be somewhat larger than the critical $U$ of graphene for the same parameter set\cite{Grphn_And_Imp}, but much less compared to $U$ in an ordinary transition metal. Hence transition material atoms, not magnetic in isolation, can become magnetic when introduced in a semi-Dirac system.
\begin{figure}[ht]
\begin{center}
\includegraphics[draft=false,bb=50 225 500 550, clip, width=\columnwidth]{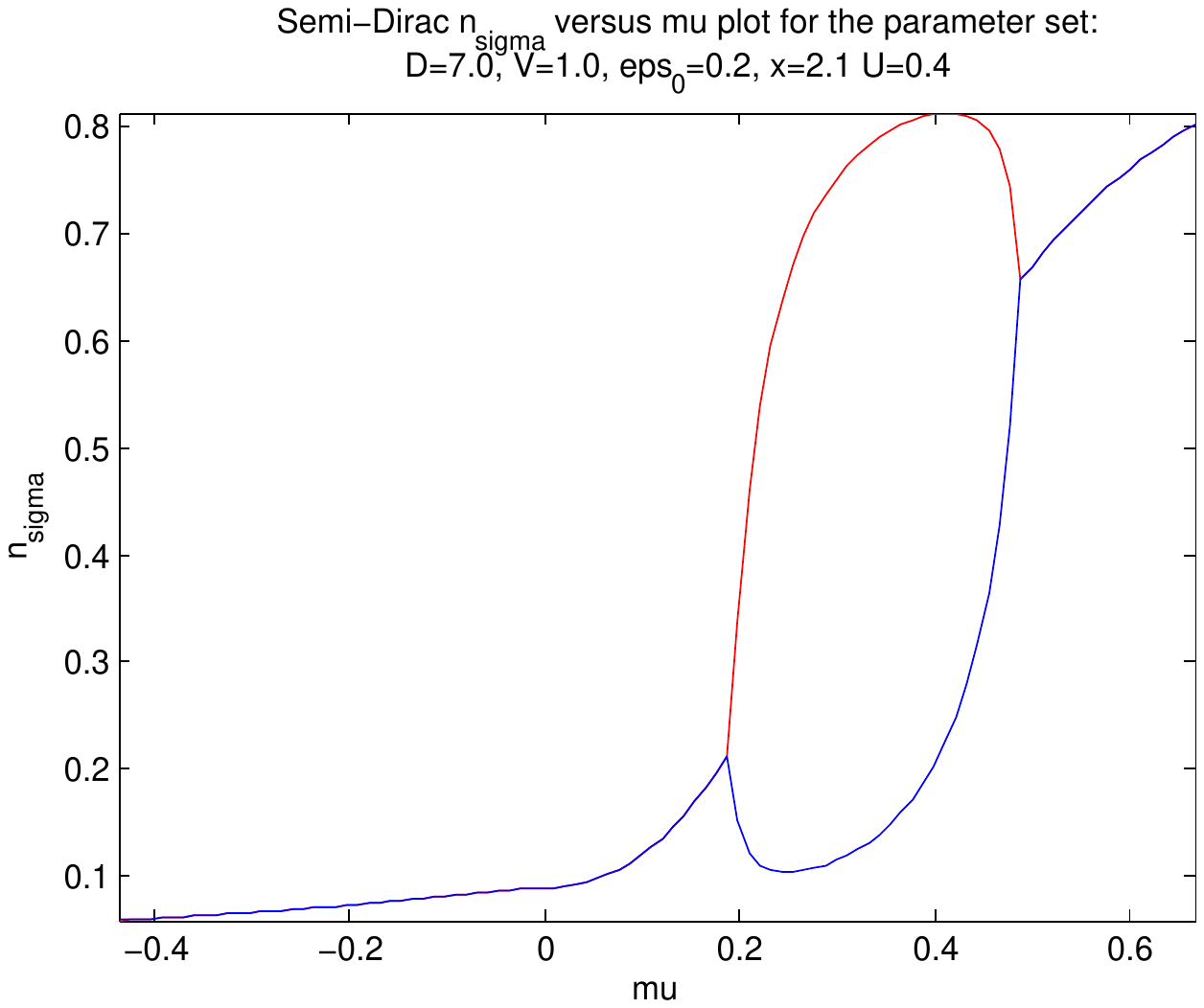}
\includegraphics[draft=false,bb=50 250 500 550, clip, width=\columnwidth]{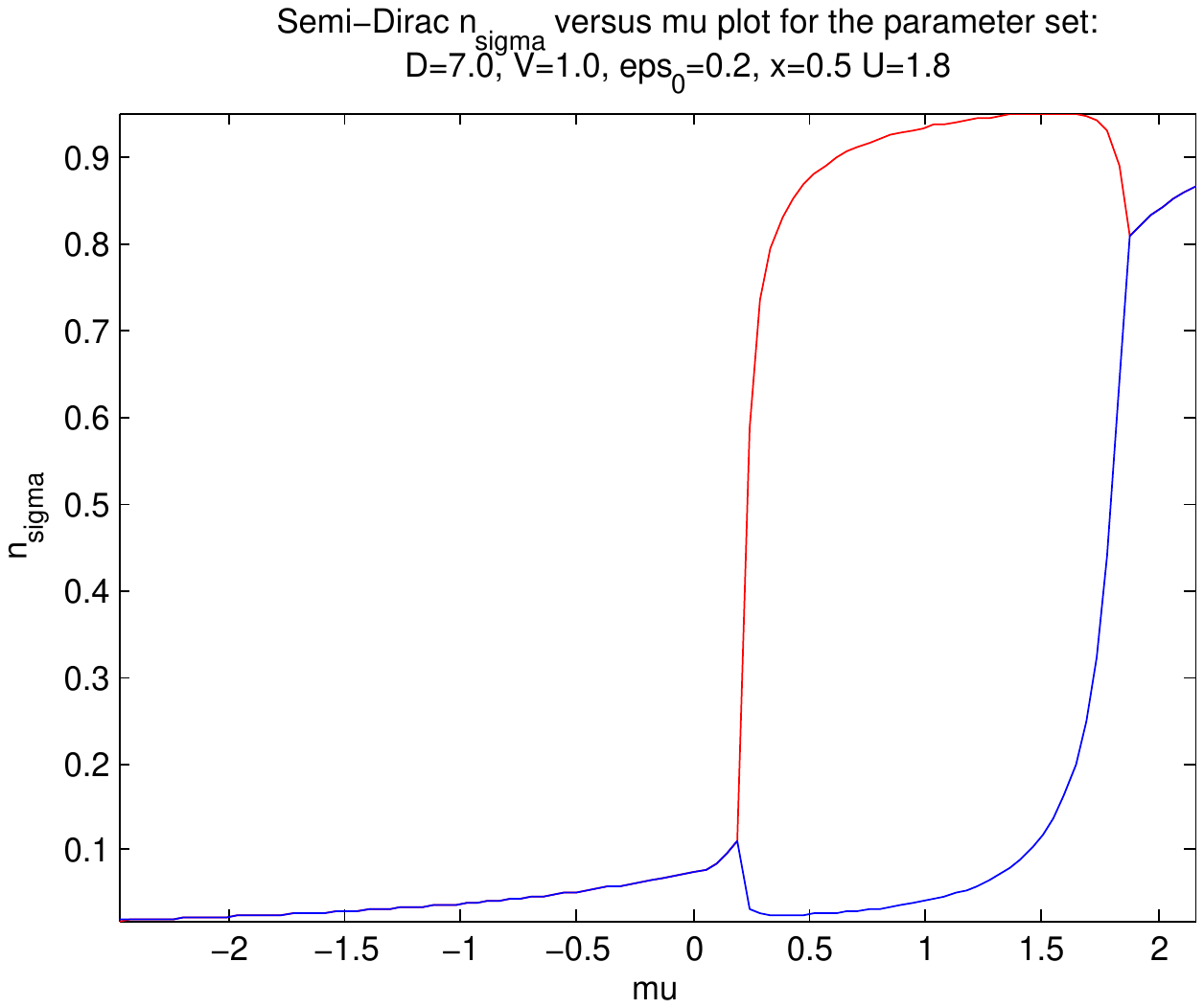}
\caption{$n_\sigma$ versus $\mu$ plot. The top and the bottom figures
have identical parameter set except for the values of $x$.}
\label{fig:n_sigma_mu}
\end{center}
\end{figure}
Next the occupations of the impurity energy level for different types of spins are plotted as functions of Fermi energy.
In Fig.\ref{fig:n_sigma_mu} the impurity occupations are plotted for for two different values of $x$'s: $2.1$ and $.5$.
Both the top and the bottom plots correspond to the the following parameter set:
$\epsilon_0=.2, V=1, \text{and} D=7$. $x=2.1$ for the top plot and $x=.5$ for the bottom one. For $x=2.1$ it is observed that there is a magnetic to non-magnetic transition \textit{above} $\mu=.2$. This is different when compared to the Dirac dispersion, for in the Dirac dispersion the transition happens at $\mu$ somewhat lesser than .2 for the same parameter set\cite{Grphn_And_Imp}. There are also other differences, like the range of $\mu$ over which there is a non-zero magnetic moment is twice for the semi-Dirac system than that in Dirac. Also, for semi-Dirac the magnetic moment is slightly larger $(\sim .65\mu_B)$ compared to Dirac's $\sim .5\mu_B$\cite{Grphn_And_Imp}for the same parameter set. The larger range of magnetism for the semi-Dirac can be attributed to its larger value of $U$ due to the appearance of $D^{3/2}$ (instead of $D^2$ as is the case for Dirac)in the denominator of the expression for $\Delta$. That means for a given set of $V$ and $D$, $\Delta$ for the semi-Dirac is larger. Hence for a given value of $x$, $U$ for the semi-Dirac is also larger as can be seen from the expression $x=\frac{\Delta D}{U}$. A large $U$ is more likely to favor magnetism, which can be seen considering the extreme case of $U \rightarrow \infty$, when due to the large energy cost of double occupation, the impurity site will always be occupied by a single spin resulting in a non-zero magnetic moment. In the same figure, at the bottom the occupation numbers are plotted for $x=.5$.
$\epsilon_0$, as well as all other parameters remain the same as in the top. Compared to the figure in the top, the non-zero magnetic moment is over a larger range of $\mu$ and the magnetic to non-magnetic transition is rather abrupt for the bottom figure. The magnetic moment in the magnetic region gets as large as $.9 \mu_B$. All these features are rather similar to the Dirac case, except for a somewhat larger range of $\mu$ for which the system stays in the magnetic phase. When $\epsilon_0$ is negative, the transition takes place for a value of $\mu$ which is barely negative (not shown in the figure). In that respect the semi-Dirac system is different from Dirac; for in Dirac, the non-magnetic to magnetic transition takes place at a larger negative value of $\mu$. The magnetic susceptibility for the impurity is computed using the following expression\cite{Grphn_And_Imp}
\begin{eqnarray}\label{eq:Chi}
\chi=-\mu_B^2\sum_\sigma \frac{dn_\sigma}{d\epsilon_\sigma}\frac{1-U\frac{dn_{-\sigma}}{d\epsilon_{-\sigma}}}
{1-U^2\frac{dn_{-\sigma}}{d\epsilon_{-\sigma}}\frac{dn_{\sigma}}{d\epsilon_{\sigma}}},
\end{eqnarray}
where $\mu_B$ is the Bohr magneton. It been checked that the susceptibilities (not shown in the plot) show sharp peaks for those values of $\mu$, at which the non-magnetic to magnetic transition takes place.

\section{conclusion}
In this paper we have discussed the Anderson's impurity problem in the context of a semi-Dirac system. The magnetic impurity
couples with the bath of electrons of the anisotropic semi-Dirac system to give rise to a unique magnetic to non-magnetic transition phenomenon.
The unconventional energy dependence of the density of states of the semi-Dirac conduction electrons results in a rather unique hybridization with the magnetic impurity atom, making it very different from either graphene or an ordinary metal. In this paper a specific model Hamiltonian for the semi-Dirac system has been used. A future direction will be to use other types of Hamiltonian for the semi-Dirac system and check the effect of magnetic impurity for that system. Semi-Dirac is a relatively new discovery, which exists in theory and in simulation. In coming years as understanding about this unique system gets better, it will be clearer if the Fermi energy can be controlled successfully by the application of some kind of a gate voltage, a direct application of which will render the semi-Dirac material valuable for spintronics. This paper is a first attempt to understand the interaction between a magnetic impurity and a specific type of semi-Dirac system, hoping that it will give an idea about the magnetic to nonmagnetic phase boundary and magnetic moment formation dependent on the parameters like impurity energy, impurity to conduction electron coupling, and Fermi energy of the system.

\end{document}